\documentclass{article}
\title{Computing Differential Equations 
       for Integrals Associated to Smooth Fano Polytopes}
\author{Hiromasa Nakayama${}^*$ 
       and Nobuki Takayama
 \footnote{Department of Mathematics, Kobe University and
    the crest Hibi project}
 }
\date{ December 24, 2010}

\usepackage{amsmath,amscd,amssymb}

\newtheorem{theorem}{Theorem}

\newtheorem{example}{Example}
\newtheorem{definition}{Definition}

\newtheorem{algorithm}{Algorithm}

\def\pd#1{ \partial_{#1} }

\def\nquad#1{\count0=0%
          \loop \ifnum#1 > \count0 \quad \advance \count0 by 1 \repeat}
\def\ord{{\mathrm ord}}

\def\p{\partial}
\def\QED{ Q.E.D. \par \bigbreak}

\begin{document}
\maketitle

Abstract: we give an approximate algorithm of computing holonomic
systems of linear differential equations for definite integrals
with parameters.
We show that this algorithm gives a correct answer in finite steps,
but we have no general stopping condition.
We apply the approximate method to find differential equations
for integrals associated to smooth Fano polytopes.
They are interested in the study of K3 surfaces and the toric mirror symmetry.
In this class of integrals, 
we can apply Stienstra's rank formula to our algorithm,
which gives a stopping condition of the approximate algorithm.

\section{Introduction}

Let $D$ be the ring of differential operators with polynomial coefficients.
A function $f(x_1, \ldots, x_n)$  is called a holonomic function
when the annihilating ideal in $D$ of $f$ is a holonomic ideal
(see, e.g., \cite[chapter 1]{SST}).
Finding systems of differential equations for the definite integral
$F(x_{m+1}, \ldots, x_n) = \int_C f(x) dx_1, \ldots, x_m$ 
where $C$ is a cycle 
is a fundamental problem in the symbolic computation.
Since the celebrating work of D.Zeilberger about 20 years ago,
a lot of algorithms have been proposed.
Among them, the algorithm given by Oaku \cite{Oakubfunc2} constructs
generators of 
$(I + \pd{1} D + \cdots + \pd{m} D ) \cap D' $
for a holonomic left ideal $I$ in $D$.
Here, $D'=K\langle x_{m+1}, \ldots, x_n,
                  \pd{m+1}, \ldots, \pd{n} \rangle$
and
$K$ is ${\bf C}$.

We give an approximate variation of this algorithm,
which improves the performance of this algorithm.
We show that the approximate algorithm gives a correct answer in finite steps,
but we have no general stopping condition.
We apply this method to find differential equations
for integrals associated to smooth Fano polytopes.
They are interested in the study of K3 surfaces and the toric mirror symmetry
\cite{Batyrev},\cite{Ishige2}, \cite{Kreuzer}, \cite{Nagano}.
For this class of integrals, 
we can apply Stienstra's rank formula \cite{Stienstra} to our algorithm,
which gives a stopping condition of the approximate algorithm.

\section{Definite Integrals associated to Smooth Fano Polytopes}

A polytope $\mathcal{P}$ is called {\it $d$-dimensional smooth Fano polytope}
when it satisfies the following $5$ conditions.
\begin{enumerate}
\item It is $d$-dimensional lattice polytope.
\item The origin is in the interior of the polytope.
\item The dual polytope $\mathcal{P}^*$ is a lattice polytope.
\item All faces are simplices ({\it simplicial polytope})
\item The vertices of each facet is a ${\bf Z}$ basis of ${\bf Z}^d$.
\end{enumerate}

A polytope is called {\it reflexive polytope} when it satisfies the conditions 1, 2, 3.

A classification of reflexive polytopes in dimensions $3$ and $4$ is
given in \cite{Kreuzer}, \cite{Kreuzer-CY}.
The following list is the list of smooth Fano polytopes 
for dimensions $2$ and $3$.
It can be obtained by extracting smooth Fano polytopes from Kreuzer's list
of reflexive Fano polytopes.
We note that \O bro(\cite{Oebro}) gave an algorithm of find all 
smooth Fano polytopes and also gave a list of them up to the dimension $7$
modulo isomorphism.
The next list is generated by the program by \O bro.

{\small
\begin{center}
\begin{tabular}{|c|c|c|}
\hline
dim & index & vertices\\
\hline 
2&0&(1,0), (0,1), (-1,-1)\\
2&1&(1,0), (0,1), (-1,0), (0,-1)\\
2&2&(1,0), (0,1), (-1,1), (0,-1)\\
2&3&(1,0), (0,1), (-1,1), (1,-1), (-1,0)\\
2&4&(1,0), (0,1), (-1,1), (1,-1), (-1,0), (0,-1)\\
\hline
3&0&(1,0,0), (0,1,0), (0,0,1), (-1,-1,-1)\\
3&1&(1,0,0), (0,1,0), (0,0,1), (-1,0,0), (0,-1,-1)\\
3&2&(1,0,0), (0,1,0), (0,0,1), (-1,-1,1), (0,0,-1)\\
3&3&(1,0,0), (0,1,0), (0,0,1), (-1,0,1), (0,-1,-1)\\
3&4&(1,0,0), (0,1,0), (0,0,1), (-1,-1,2), (0,0,-1)\\
3&5&(1,0,0), (0,1,0), (0,0,1), (-1,-1,2), (0,1,-1), (0,0,-1)\\
3&6&(1,0,0), (0,1,0), (0,0,1), (-1,0,1), (1,0,-1), (-1,-1,0)\\
3&7&(1,0,0), (0,1,0), (0,0,1), (-1,0,1), (0,-1,0), (0,0,-1)\\
3&8&(1,0,0), (0,1,0), (0,0,1), (-1,0,1), (0,0,-1), (0,-1,-1)\\
3&9&(1,0,0), (0,1,0), (0,0,1), (-1,0,0), (0,-1,0), (0,0,-1)\\
3&10&(1,0,0), (0,1,0), (0,0,1), (-1,0,1), (0,-1,1), (0,0,-1)\\
3&11&(1,0,0), (0,1,0), (0,0,1), (-1,0,1), (0,1,-1), (0,-1,0)\\
3&12&(1,0,0), (0,1,0), (0,0,1), (-1,0,1), (0,-1,1), (0,1,-1), (0,-1,0)\\
3&13&(1,0,0), (0,1,0), (0,0,1), (-1,0,1), (0,-1,1), (0,1,-1), (0,0,-1)\\
3&14&(1,0,0), (0,1,0), (0,0,1), (-1,0,1), (0,1,-1), (0,-1,0), (0,0,-1)\\
3&15&(1,0,0), (0,1,0), (0,0,1), (-1,0,1), (1,0,-1), (-1,0,0), (0,-1,0)\\
3&16&(1,0,0), (0,1,0), (0,0,1), (-1,0,1), (0,-1,1), (0,1,-1), (0,-1,0), (0,0,-1)\\
3&17&(1,0,0), (0,1,0), (0,0,1), (-1,0,1), (1,0,-1), (-1,0,0), (0,-1,0), (0,0,-1) \\
\hline
\end{tabular}
\end{center}
}

Let $\{a_1, \cdots, a_m\}$ be the vertices of a $n$-dimensional smooth Fano polytope $P_{n,k}$
and $a_{m+1}$  the origin.
The suffix $k$ is the index in the list of the smooth Fano polytopes.
We put
$ f_{n,k}(x,t) = \sum_{i=1}^{m+1} x_i t^{a_i} $.
We are interested in a definite integral with parameters associated to these polytopes
defined by
$$ F_{n,k}(x) = \int_C f_{n,k}(x,t)^{-1} t_1^{-1} \cdots t_n^{-1} dt_1 \cdots dt_n $$
where $x$ is a generic parameter vector and
$C$ is a cycle in $H_n(T_x,{\bf C})$,
$T_x = \{t \in ({\bf C}^*)^n ~|~ f(t,x) \neq 0\}$.
The definite integral is called 
a {\it period integral associated to} $P_{n,k}$.
The function $F_{n,k}$ satisfies $A$-hypergeometric system
for $A=(a_1, \ldots, a_{n+1})$ 
and $\beta=(-1, 0, \ldots, 0)^T$
and the system is reducible
\cite{Batyrev}.
The construction of the subsystem is implicit in \cite{Batyrev}.
We are interested in an explicit expression of the subsystem
of the $A$-hypergeometric system satisfied by $F_{n,k}$.

\begin{example} \rm
The vertices of $P_{3,0}$ are
 $\{(1,0,0), (0,1,0), (0,0,1), (-1,-1,-1)\}$.
 Then, we have, by definition,
 $$F_{3,0}(x) = \int_C (x_1 t_1 + x_2 t_2 + x_3 t_3 + x_4 t_1^{-1} t_2^{-1} t_3^{-1})^{-1}
t_1^{-1} t_2^{-1} t_3^{-1} dt_1 dt_2 dt_3,
$$
which is nothing but the ${\cal A}$-hypergeometric integral
for the matrix
$$ A=
\begin{pmatrix}
1 & 1 & 1 & 1 & 1 \\
1 & 0 & 0 & -1& 0 \\
0 & 1 & 0 & -1& 0 \\
0 & 0 & 1 & -1& 0 \\
\end{pmatrix}
$$  
and  $\beta = (-1,0,0,0)^{T}$.
It satisfies the ${\cal A}$-hypergeometric system $H_A(\beta)$ of which
holonomic rank is $4$.
As we will see later, the function $F_{3,0}(x)$ satisfies a system of which rank is $3$.
\end{example}

We note that the problem of finding systems of differential equations for period integrals
associated to $P_{n,k}$ has been considered to study 
the moduli space of the family of  hypersurfaces $f(t,x)=0$.
Ishige \cite{Ishige} studied the case of 
$P_3=\{(1,0,0),(0,1,0),(0,0,1),(0,0,-1),(-1,-1,-1)\}$
and
Nagano \cite{Nagano} studied the cases of
\begin{align*} 
 P_2&=\{(1,0,0),(0,1,0),(0,0,1),(-1,-1,0),(0,0,-1)\}, \\
 P_4&=\{(1,0,0),(0,1,0),(0,0,1),(0,0,-1),(-1,-1,-2)\}, \\
 P_5&=\{(1,0,0),(0,1,0),(0,0,1),(-1,-1,0),(-1,-1,-1)\}, \\
 P_r&=\{(1,0,0),(0,1,0),(0,0,1),(0,-1,-1),(-1,0,-1)\}
\end{align*}
The correspondence of our table and Nagano's table is as follows. \\
\begin{tabular}{|c||c|c|c|c|} \hline
$P_i$ :(dim,index) &
$P_2$ : (3,1) &
$P_3$ : (3,2) &
$P_4$ : (3,4) &
$P_5$ : (3,3) \\ 
\hline
\end{tabular} \\
$P_r$ is not simplicial polytope.
Nagano gives series expansion of the function $F_{n,k}$ and
gives a heuristic method to find annihilating operators of the function
by the method of undetermined coefficients
and this method is enough for his purpose.

We will give an efficient algorithm to derive systems of differential equations
for period integrals in this paper.
Our method gives a basis of associated cohomology groups.

\section{Computational Bottleneck of Integration Algorithm for $D$-modules}

Computation of differential equations satisfied by a definite integral with parameters
can be reduced to the $D$-module theoretic integral of the annihilating ideal
of the integrand
(\cite{Oakubfunc2}, \cite{OT}, \cite{SST}).
This method works for any integrand which is a holonomic function.
In particular, we can apply it to find differential equations
for the function $F_{n,k}$.
However, it requires huge computational resources in general.
In this section, we will explain what are bottlenecks of this algorithm in case of
our problem.

We denote by $D$ the ring of differential operators of polynomial coefficients
in $n$ variables
$$ D = K \langle x_1, \ldots, x_m, x_{m+1}, \ldots x_n, \pd{1}, \ldots, \pd{m},
\pd{m+1}, \ldots, \pd{n} \rangle $$
and by $D'$ that in $n-m$ variables
$$ D' = K \langle x_{m+1}, \ldots, x_n, \pd{m+1}, \ldots, \pd{n} \rangle$$
Let $I$ be a left ideal of $D$.
The integration ideal of $I$ with respect to $x_1, \cdots, x_m$
is 
$$ J = (I + \pd{1} D + \cdots + \pd{m} D) \cap D' , $$
which is known to be a holonomic left ideal of $D'$.

When the ideal $I$ is the annihilating ideal of a function $f(x_1, \ldots, x_n)$,
the integration ideal $J$ annihilates the definite integral
$$\int_C f(x_1, \cdots, x_m, x_{m+1}, \cdots, x_n) dx_{1} \cdots dx_m$$ 
for any $m$-cycle $C$
(see, e.g., \cite[Chap5 ]{SST}).

Let us review the $D$-module theoretic integration algorithm
following Oaku \cite{Oakubfunc2}
to show bottlenecks of this algorithm.
We define the ring isomorphism of $D$ $\mathcal{F}$ by 
$$\mathcal{F}(x_i) = 
\begin{cases} 
-\pd{i} ~~~(1 \leq i \leq m) \\
x_i     ~~~(m < i \leq n)
\end{cases}
,  
\mathcal{F}(\pd{i}) = 
\begin{cases} 
x_i     ~~~(1 \leq i \leq m) \\
\pd{i}  ~~~(m < i \leq n)
\end{cases}
$$ 
and we call it the {\it Fourier transformation}.
The inverse map of $\mathcal{F}$ is called the {\it inverse Fourier transformation}
and denoted by $\mathcal{F}^{-1}$.
We use the symbol ${\rm in}_{(-w,w)}$ defined in \cite{SST}
to express the initial ideal.
The generic $b$-function of $I$ with respect to the weight $w \in {\bf R}^n$ 
is the monic generator of the ideal
$ {\rm in}_{(-w,w)}(I) \cap {\bf C}[s] $
in ${\bf C}[s]$ where
$s = \sum_{i=1}^n w_i x_i \pd{i}$.

\begin{algorithm} 
[Computation of integration ideal, \cite{Oakubfunc2}, \cite{OT}, \cite{SST}]
\label{alg:int-ideal}
~~~ 
\begin{itemize}
\item[Input:] A set of generators of holonomic left $D$ ideal $I$.  \\
A weight vector $w = (w_1, \ldots, w_m, w_{m+1}, \ldots, w_n)$  satisfying\\
 $w_1, \ldots, w_m > 0, w_{m+1} = \cdots = w_n = 0$.
\item[Output:]  A set of generators of the integration ideal of $I$ with respect to $x_1, \ldots, x_m$.
\end{itemize}

\begin{enumerate}
\item  Compute the restriction ideal of $\mathcal{F}(I)$ with respect to the weight vector  $w$.
This step consists of the following steps.
      \begin{enumerate}
      \item Compute a Gr\"obner basis $ G=\{h_1, \cdots, h_l\}$ of $\mathcal{F}(I)$ with respect to
      the order $<_{(-w,w)}$.
      \item Compute the generic $b$-function $b(s)$ of $\mathcal{F}(I)$
       with respect to $(-w,w)$.
       If there is no non-negative integral root of $b(s)=0$, then output $1$.
       Let $s_0$ be the maximal non-negative integral root of $b(s)=0$.
      \item Put $m_i = \ord_{(-w,w)}(h_i)$, \\ 
            $\mathcal{B}_d = \{\pd{1}^{i_1} \cdots \pd{m}^{i_m} ~|~
            i_1 w_1 + \cdots + i_m w_m \leq d \}$, \\
            $r = \#\{(i_1, \ldots, i_m) ~|~
            i_1 w_1 + \cdots + i_m w_m \leq s_0\}$.
      \item Put $\mathcal{B}=\{(\p^{\beta} h_i)|_{x_1=\cdots=x_m=0}
            \mid \p^{\beta} \in \mathcal{B}_{s_0-m_i}, h_i \in G \}$.
            Here, $(\p^{\beta} h_i)|_{x_1 = \cdots = x_m = 0}$  is calculated by substituting
            $x_1 = \cdots = x_m =0$ after expanding 
            $\p^{\beta} h_i$ into the form $\sum_{u,v} c_{u,v} x^u \p^v (c_{u,v} \text{ is a constant})$.
      \end{enumerate}
\item 
Elements in
$\mathcal{F}^{-1}(\mathcal{B})$
can be expressed as
$$ \sum_{w_1 i_1+\cdots+w_m i_m \leq d} c_{i_1, \cdots, i_m} x_1^{i_1} \cdots x_m^{i_m} ~ 
(c_{i_1, \cdots, i_m} \in D'). $$
We regard it as an element of $(D')^r$.
In other words, we consider a vector  
$(c_{i_1, \cdots, i_m})_{w_1 i_1 + \cdots + w_m i_m \leq d}$
of which entries belong to $D'$.
Let $M$ be the left $D'$ module generated by them.
\item 
Compute a Gr\"obner basis of $M$ with respect to an POT order such that 
the element standing for $x_1^0 \cdots x_m^0$ is the minimum.
Collect elements in the Gr\"obner basis such that  all entries 
except the entry standing for 
$x_1^0 \cdots x_m^0$ is $0$.
We regard these elements as elements of $D'$.
Output them.
\end{enumerate}
\end{algorithm}

\begin{example}
[Period integral associated to a $2$ dimensional smooth Fano polytope]
~~~

We will derive a system of differential equations for the period integral
$$F_{2,0}(x) = \int_C (x_1 t_1 + x_2 t_2 + x_3 t_1^{-1} t_2^{-1} + x_4)^{-1} t_1^{-1} t_2^{-1} dt_1 dt_2 $$
associated to the smooth Fano polytope $P_{2,0}$
by the algorithm we have presented.
\begin{enumerate}
\item 
Obtain the annihilating ideal $I$ of 
$$f_{2,0}(x,t) = (x_1 t_1 + x_2 t_2 + x_3 t_1^{-1} t_2^{-1})^{-1} t_1^{-1} t_2^{-1}$$
by Oaku's algorithm (\cite{Oakubfunc1}).


\item
For the ideal $I$,
we compute the integration ideal of $I$ with respect to $t_1, t_2$
$$J = (I + \partial_{t_1} D + \partial_{t_2} D) \cap D'$$
Here,
$ D = K \langle x_1,x_2,x_3,x_4,t_1,t_2, \p_{x_1},\p_{x_2},\p_{x_3},\p_{x_4},\p_{t_1},\p_{t_2} \rangle $, \\
$ D' = K \langle x_1,x_2,x_3,x_4,\p_{x_1},\p_{x_2},\p_{x_3},\p_{x_4} \rangle$.

\end{enumerate}
Generators of $J$ obtained by Algorithm \ref{alg:int-ideal} are
{\small
\begin{align*}
&(x_4^3+27 x_1 x_2 x_3) \p_{x_4}^2+3 x_4^2 \p_{x_4}+x_4,~
9 x_2 x_3 \p_{x_4}^2-x_4^2 \p_{x_1} \p_{x_4}-x_4 \p_{x_1},~
9 x_1 x_3 \p_{x_4}^2-x_4^2 \p_{x_2} \p_{x_4}-x_4 \p_{x_2}, \\
&-9 x_1 x_2 \p_{x_4}^2+x_4^2 \p_{x_3} \p_{x_4}+x_4 \p_{x_3},~
-3 x_3 \p_{x_4}^2-x_4 \p_{x_1} \p_{x_2},~
-3 x_2 \p_{x_4}^2-x_4 \p_{x_1} \p_{x_3},~
-3 x_1 \p_{x_4}^2-x_4 \p_{x_2} \p_{x_3}, \\
&-\p_{x_4}^3+\p_{x_1} \p_{x_2} \p_{x_3},~ 
x_4 \p_{x_4}+3 x_1 \p_{x_1}+1,~
-x_4 \p_{x_4}-3 x_2 \p_{x_2}-1,~
x_4 \p_{x_4}+3 x_3 \p_{x_3}+1
\end{align*}
}
The four elements in the last line are elements in 
the $A$-hypergeometric ideal $H_A(\beta)$
and other elements are not.
Here, we put $A=\begin{pmatrix} 1&1&1&1\\1&0&-1&0\\0&1&-1&0 \end{pmatrix}, \beta=(-1,0)^{T}$.

\end{example}

We present timing data obtained by applying these algorithms
to our problem.
The timing data are taken on Risa/Asir (\cite{Asir}) version 20100206
and the library package {\tt nk\_restriction.rr}
running on a machine with Xeon X5470 (3.33GHz) CPU and 3.6 GB memory
except the computation of annihilating ideal,
which is computed on Singular and the same machine.
It is known that the Singular implementation of it is the best one among several 
implementations.
In the table below,
the field Ann stands for Brian\c{o}n-Maisonobe algorithm of computing annihilating ideal
(\cite{Briancon}, \cite{Levandovskyy})
implemented in Singular,
the field generic-b for the steps 1 (a), (b) of  Algorithm \ref{alg:int-ideal},
the field base for the steps 1 (c), (d) and the step 2 
of  Algorithm \ref{alg:int-ideal},
and the field GB for the step 3 of Algorithm \ref{alg:int-ideal}.
Each entry means seconds. 
The symbol ``---'' means that we could not obtain the result
in one day.
\begin{center}
\begin{tabular}{|c|c|c|c|c|c|}
\hline
dim & index & Ann   & $<_{(-w,w)}$-GB and generic-b & base & GB    \\
\hline
2   & 0     & $<1$    & 0.004     & 0.04 & 0.004 \\
2   & 1     & $<1$    & 0.022     & 0.04 & 0.008 \\
2   & 2     & $<1$    & 0.035     & 0.03 & 0.019 \\
2   & 3     & 2       & 0.58      & 0.11 & 0.19  \\
2   & 4     & 180     & 93        & 1.6  & 14    \\
\hline
3   & 0     & 1     & 0.052     & 0.048& 0.020 \\
3   & 1     & 1     & 0.20      & 0.09 & 0.052 \\
3   & 2     & 3     & 0.36      & 0.16 & 0.10  \\
3   & 3     & 5     & 1.6       & 0.25 & 0.14  \\
3   & 4     & 14    & 1.4       & 0.25 & 0.013 \\
3   & 5     & 4165  & 1710      & 726  & 5726  \\
3   & 6     & 2383  & 5037      & 1007 & 6194   \\
3   & 7     & 135   & 92        & 87   & 257   \\
3   & 8     & 1551  & 1618      & 499  & 2227  \\
3   & 9     & 16    & 13        & 34   & 86    \\
3   & 10    & 1492  & 852       & 183  & 994   \\
3   & 11    & 6213  & 1588      & 406  & 1839  \\
3   & 12    & ---   & ---       & ---  & ---   \\
\multicolumn{6}{c}{$\cdots \cdots \cdots$ } \\ 
3   & 17    & ---   & ---       & --- & ---  \\
\hline
\end{tabular}
\end{center}

The data tell us that if we can accelerate
the steps ``Ann''(computation of annihilating ideal) 
and ``$<_{(-w,w)}$-GB and generic-b'' (computation of $(-w,w)$-Gr\"obner basis and generic $b$-function),
then we may be able to speed up the computation in this list.
In the next section, we will suggest a new efficient method. 

\section{Approximate Integration Algorithm}

\begin{definition}[Approximate annihilating ideal]~~~ \\
\rm Let $f, g$ be polynomials.
The $i$-th {\it approximate annihilating ideal} of the rational function
$\frac{f}{g}$
is the ideal generated by the elements of which $({\bf 0},{\bf 1})$-order 
is less than or equal
to $i$ in ${\rm Ann}_D \frac{f}{g}$.
We denote it by
${\rm Ann}^{(i)}_D \frac{f}{g}$.
\end{definition}

The following method is simple, but is very useful.
\begin{algorithm}
[Computation of approximate annihilating ideal \cite{CastroUcha}]~~~ 
\label{alg:app-ann} \rm

Input:  rational function $\frac{f}{g}$, order $i$ 

Output : generators of $i$-th approximate annihilating ideal 
${\rm Ann}_D^{(i)} \frac{f}{g}$.

\begin{enumerate}
\item 
Put $P = \sum a_\alpha \partial^\alpha$
where $a_\alpha$ are polynomials to be determined
and the sum runs over all $\alpha$ such that
the order of 
$\partial^{\alpha} ~~~ (\alpha_1 + \cdots + \alpha_n \leq i)$ 
is less than or equal to $i$.
\item 
Apply the differential operator $P$ to 
the rational function 
$\frac{f}{g}$
and derive a relation for polynomials $a_\alpha$ 
so that $P \bullet \frac{f}{g} = 0$.
The polynomials $a_\alpha$ can be determined by a syzygy computation
in the ring of polynomials.
\end{enumerate}
\end{algorithm}

\begin{example}
[Computation of approximate annihilating ideal]
~~~  \rm

We will compute the first approximate annihilating ideal for
$$f_{2,0}(t,x) = \frac{1}{x_1 t_1^2 t_2 + x_2 t_1 t_2^2 + x_3 + x_4 t_1 t_2}.$$
\begin{enumerate}
\item 
Apply 
$P=a_1 \p_{t_1} + a_2 \p_{t_2} + a_3 \p_{x_1} + a_4 \p_{x_2} + a_5 \p_{x_3} + a_6 \p_{x_4} + a_7$
($a_i$ are polynomials to be determined) to 
$f_{2,0}(t,x)$. 
Then, the numerator $n(t,x)$ is
\begin{align*}
n(t,x) =& (-t_2 x_4-t_2^2 x_2-2 t_1 t_2 x_1) a_1+(-t_1 x_4-2 t_1 t_2 x_2-t_1^2 x_1) a_2 - \\
        & t_1^2 t_2 a_3-t_1 t_2^2 a_4-a_5-t_1 t_2 a_6+(t_1 t_2 x_4+x_3+t_1 t_2^2 x_2+t_1^2 t_2 x_1) a_7
\end{align*}
\item 
We want to determine $a_i$ so that $n(t,x) = 0$.
Let $c_i$ be the coefficient of $a_i$ of $n(t,x)$.
Any element of the syzygy module ${\rm Syz}(c_1, \cdots, c_7)$ 
gives $\{ a_i \}$ such 
that $n(t,x)=0$ and if a set $\{ a_i \}$ makes $n(t,x)=0$,
then it is an element of the syzygy.
The syzygy can be obtained by Gr\"obner basis computation in the 
ring of polynomials and it is generated by
\begin{align*}
&(-1,0,0,0,t_2 x_4+t_2^2 x_2,2 x_1,0),(0,-1,0,0,t_1 x_4+t_1^2 x_1,2 x_2,0),(0,0,-1,0,0,t_1,0), \\
&(0,0,0,-1,0,t_2,0),(0,0,0,0,-t_1 t_2,1,0),(0,0,0,0,x_3,x_4+t_2 x_2+t_1 x_1,1), \\
&(0,0,0,0,0,t_1 t_2 x_4+x_3+t_1 t_2^2 x_2+t_1^2 t_2 x_1,t_1 t_2)
\end{align*}
The set of differential operators standing for them
\begin{align*}
&-\p_{t_1}+(t_2 x_4+t_2^2 x_2) \p_{x_3}+2 x_1 \p_{x_4},-\p_{t_2}+(t_1 x_4+t_1^2 x_1) \p_{x_3}+2 x_2 \p_{x_4},-\p_{x_1}+t_1 \p_{x_4}, \\
&-\p_{x_2}+t_2 \p_{x_4},-t_1 t_2 \p_{x_3}+\p_{x_4},x_3 \p_{x_3}+(x_4+t_2 x_2+t_1 x_1) \p_{x_4}+1, \\
&(t_1 t_2 x_4+x_3+t_1 t_2^2 x_2+t_1^2 t_2 x_1) \p_{x_4}+t_1 t_2 
\end{align*}
is a set of generators of the first approximate annihilating ideal.
\end{enumerate}
\end{example}

It is an interesting open question that at which order
the approximate annihilating ideal equals to the annihilating ideal.
Castro and Ucha \cite{CastroUcha} proved that
if a polynomial $f$ in two variables is weighted homogeneous,
then we have
${\rm Ann}_D^{(1)}\frac{1}{f} = {\rm Ann}_D \frac{1}{f}$.

Utilizing approximate annihilating ideal, we give 
an approximate integration algorithm.

The syzygy computation in the ring of polynomials
is faster than computation of the annihilating ideal in $D$ in general.
Then, we can expect that our approximate integration algorithm runs
faster than the standard one.
However, we need a stopping condition to stop the approximate  procedure.

\begin{algorithm}
[Approximate integration algorithm]
~~~ \rm
\begin{itemize}
\item[Input :] $\frac{f}{g}$ is a rational function. $m$ is a natural number.
\item[Output :] A subideal $J'$ 
of the integration ideal of ${\rm Ann}_D \frac{f}{g}$.
\end{itemize}
\begin{enumerate}
\item  Compute $J = {\rm Ann}_D^{(m)} \frac{f}{g}$ (Algorithm \ref{alg:app-ann})
\item  If $J$ is holonomic, then apply the ($D$-module) theoretic integration algorithm to $J$ and put it $J'$, else put $J'=\langle 0 \rangle$. 
\end{enumerate}
\end{algorithm}

\begin{theorem}
There exists a natural number $m$ such that
the approximate integration ideal $J'$ agrees with
the integration ideal. 
\end{theorem}

{\it Proof}.
Since $D$ is a Noetherian ring, there exists $m$ such that
${\rm Ann}_D^{(m)} \frac{f}{g} = {\rm Ann}_D \frac{f}{g}$.
The theorem follows from this fact.
\QED

\begin{example}
[$m$ is $2$.]
~~~  \rm

We use the example by Castro and Ucha(\cite{CastroUcha}).
Consider the polynomial $f = x^4+y^5+xy^4$ (Reiffen curves).
We have
$$ {\rm Ann}_D^{(1)}\frac{1}{f} \subsetneq {\rm Ann}_D^{(2)}\frac{1}{f} = 
   {\rm Ann}_D \frac{1}{f} $$
%

The first approximate integration ideal for $\frac{1}{f}$ with respect to $x$
is 
$J^{(1)}= D\cdot\{y P\}$
and the integration ideal $J$ is
$ D\cdot\{P\}$.
Here,
$$P = (-27 y^4+256 y^3) \p_y^3+(-432 y^3+3456 y^2) \p_y^2+(-1896 y^2+12336 y) \p_y-2184 y+10920$$
For  $m=2$, the approximate integration ideal $J^{(2)}$
agrees with  $J$.
\end{example}

Let us come back to our problem of computing differential equations
for period maps associated to smooth Fano polytopes.
In application, we do not need to find the exact integration ideal
and we may find an approximate integration ideal of which holonomic rank
agrees with that of the exact solution. 
For this purpose, the method given by Stienstra gives a way to find
the approximate order ``m''.
We follow the notation of the paper of Stienstra \cite[p.434]{Stienstra}

${\bf T} \setminus {\bf Z}_s$ is defined as
${\bf C}^{n-1} \setminus \{ x_1 \cdots x_{n-1} \cdot f(x_1, \cdots, x_{n-1})=0 \}$
where $f(x_1, \cdots, x_{n-1})$ is defined as follows from the set of points 
$A = 
\begin{pmatrix}
1   & 1   & \cdots & 1 \\
a_1 & a_2 & \cdots & a_l
\end{pmatrix}
$.
$$ f(x_1, \cdots, x_{n-1}) = \sum_{i=1}^l u_i x^{a_i} ~~~ (u_i \in {\bf C}) $$

\begin{algorithm} \label{algorithm:approxint}
(An algorithm finding a set of differential operators
for integrals associated to a class of smooth Fano polytopes) 
\begin{enumerate}
\item Evaluate the dimension 
$ r = {\rm dim}\, W_n H^{n-1}({\bf T} \setminus {\bf Z}_s) $
by Stienstra's method (\cite[p435 (57), p448 Th.10 (iv)]{Stienstra}).
\item $i=1$.
\item Compute the integral ideal $J$ for ${\rm Ann}_D^{(i)}(I)$.
\item If the holonomic rank $J$ is equal to $r$, then stop and output $J$,
else $i \leftarrow i+1$ and goto the step 3.
\end{enumerate}
\end{algorithm}

\begin{theorem} 
~~~ 

\begin{enumerate}
\item 
Suppose that $A$ is smooth Fano (reflexive Gorenstein) and 
admits a unimodular regular triangulation and, moreover,
\begin{equation} \label{eq:weighteq}
 r = {\rm dim}\, W_n H^{n-1}({\bf T} \setminus {\bf Z}_s) 
   = {\rm dim}\,  H^{n-1}({\bf T} \setminus {\bf Z}_s)
\end{equation}
Then, 
Algorithm \ref{algorithm:approxint} outputs a set of differential operators
of which holonomic rank agrees with $r$.
\item 
If Algorithm \ref{algorithm:approxint} stops, 
then it implies the equality(\ref{eq:weighteq}).
\end{enumerate}
\end{theorem}

{\it Proof}.
(1)
Since the annihilating ideal $J={\rm Ann}_D\, (I)$ is finitely generated,
there exists $i$ such that
$D \cdot {\rm Ann}_D^{(i)}(I) = {\rm Ann}_D \, (I)$.
The integration module of $D/J$ is isomorphic to 
$H^{n-1}({\bf T} \setminus {\bf Z}_s)$ \cite{OT}.
Therefore, our algorithm outputs the integration ideal of which 
rank is equal to $r$.
(2) We have
$W_{n-1} H^{n-1}({\bf T} \setminus {\bf Z}_s) \subseteq 
 W_n H^{n-1}({\bf T} \setminus {\bf Z}_s) \subseteq 
  H^{n-1}({\bf T} \setminus {\bf Z}_s)$
in general.
The claim (2) follows from this inclusion.
\QED

\begin{example} \rm
We consider the case $P_{2,0}$.
Vertices of the polytope $P_{2,0}$ are 
$a_1 = (1,0), a_2 = (0,1), a_3 = (-1,-1)$.
We put $a_4 = (0,0)$ and
$$ A=\begin{pmatrix} 1&1&1&1 \\ 1&0&-1&0 \\ 0&1&-1&0 \end{pmatrix}=(a_{ij}) $$
Put 
$\mathcal{T} = \{1,2,3,4,12,13,23,124,234,134\}$.
Here, $i_1 \cdots i_s$ denotes the simplex spanned by
$a_{i_1}, \cdots, a_{i_s}$.
Then, $\mathcal{T}$ is a unimodular triangulation
of the set of points $\{ a_1, \ldots, a_4 \}$.
Following Stienstra \cite{Stienstra}, we consider the ring
$\mathcal{R}_{A,\mathcal{T}} = {\bf Z}[c_1, c_2, c_3, c_4]/\mathcal{J}$
where $c_i$  is a variable standing for the point $a_i$
and
the ideal $\mathcal{J}$ of the ring 
${\bf Z}[c_1, c_2, c_3, c_4]$ is 
$$ \mathcal{J} = \langle c_1 + c_2 + c_3 + c_4, c_1 - c_3, c_2 - c_3, c_1 c_2 c_3 \rangle $$
The first $3$ linear polynomial stand for $A$ and the last monomial
stands for a simplex which is not in the triangulation $\mathcal{T}$.
Put
$core \mathcal{T} = \{ \mbox{intersection of the maximal simplices} \}$
and
$c_{core} = \prod_{ i \in core \mathcal{T}} c_i$.
In this example,
$c_{core} = c_4$.
Stienstra's theorem claims that the rank of 
$\mathcal{R}_{A,\mathcal{T}}/{\rm Ann}_{\mathcal{R}_{A,\mathcal{T}}} c_{core}$
as ${\bf Z}$-module gives a lower bound of the holonomic rank
of the approximate integration ideal.
The first approximate integration ideal gives this lower bound 
for this example and then we obtain the answer.
\end{example}

We have applied this algorithm  and obtain the following two tables.
The field ``tank of $H_A(\beta)$'' is the holonomic rank of the hypergeometric
system $H_A(\beta)$ and the field ``rank of $W_n H^{n-1}$'' is the lower bound
evaluated by Stienstra's method.
The field  ``AppAnn'' in the second table is the time to find
the first order approximate annihilating ideal.
It is an interesting  observation that we need only first order operators.

\begin{center}
\begin{tabular}{|c|c|c|c|c|c|}
\hline
dim & index & rank of $H_A(\beta)$ &  rank of $W_n H^{n-1}$ \\
\hline
2   &     0 &          3   &            2 \\
2   &     1 &          4   &            2 \\
2   &     2 &          4   &            2 \\
2   &     3 &          5   &            2 \\
2   &     4 &          6   &            2 \\
\hline
3   &     0 &          4   &            3 \\
3   &     1 &          6   &            4 \\
3   &     2 &          6   &            4 \\
3   &     3 &          6   &            4 \\
3   &     4 &          6   &            4 \\
3   &     5 &          8   &            5 \\
3   &     6 &          8   &            5 \\
3   &     7 &          8   &            5 \\
3   &     8 &          8   &            5 \\
3   &     9 &          8   &            5 \\
3   &    10 &          8   &            5 \\
3   &    11 &          8   &            5 \\
3   &    12 &         10   &            6 \\
3   &    13 &         10   &            6 \\
3   &    14 &         10   &            6 \\
3   &    15 &         10   &            6 \\
3   &    16 &         12   &            7 \\
3   &    17 &         12   &            7 \\
\hline
\end{tabular}
\end{center}

\begin{center}
\begin{tabular}{|c|c|c|c|c|c|}
\hline
dim & index & AppAnn& $<_{(-w,w)}$-GB and generic-b & base & GB    \\
\hline
3   & 5     & 0.12  & 1253      & 10   & 5.8   \\
3   & 6     & 0.18  & 11781     & 2130 & 67    \\
3   & 7     & 0.18  & 32        & 2.3  & 0.71  \\
3   & 8     & 0.18  & 4014      & 69   & 8.7   \\
3   & 9     & 0.18  & 9.1       & 1.5  & 0.34  \\
3   & 10    & 0.18  & 248       & 3.4  & 1.8   \\
3   & 11    & 0.21  & 572       & 6.9  & 4.3   \\
3   & 12    & ---   & ---       & ---  & ---   \\
\multicolumn{6}{c}{$\cdots \cdots \cdots$ } \\ 
3   & 17    & ---   & ---       & --- & ---  \\
\hline
\end{tabular}
\end{center}

\section{Conclusion}

We have succeeded to efficiently find differential equations
for period maps standing for $P_{3,5}$ to $P_{3,11}$
by our new approximate method.
Explicit expressions of differential equations for period maps
and programs used in this paper can be found in
{\tt http://www.math.kobe-u.ac.jp/OpenXM/Math/sfano}

Let us briefly discuss two future research plans of this work.

The first one is to utilize the construction of all irreducible quotients
of $A$-hypergeometric $D$-module by Saito \cite{saito}.
To do this, we need to give a new integration algorithm to fit to his
construction.  

The second one is to improve the final step 
of computing a $<_{(-w,w)}$-Gr\"obner basis.
Let $R$ be the ring of differential operators with rational function
coefficients with the variables
$x_1, \ldots, x_m$, $x_{m+1}, \ldots, x_n$
and $R'$ the ring of differential operators with the variables
$x_{m+1}, \ldots, x_n$.
Chyzak \cite{chyzak} gives an approximate algorithm to find elements of 
$(RI + \sum_{i=1}^m \pd{i} R  ) \cap R'$
by utilizing the method of undermined coefficients.
Since we have to find an approximate basis of the cohomology group
to use Stienstra's criterion,
we cannot use the $R$ and we need to use $D$.
By this reason, we cannot try this idea with the current implementation
{\tt mgfun}, which uses $R$, of this algorithm in the Maple.
However, Chyzak's algorithm itself can be easily modified for $D$.
It is our future project to give an efficient implementation of this 
approximate algorithm for $D$
to apply to our problem.

\end{document}